# Fast characterization of optically detected magnetic resonance spectra via data clustering

Dylan G. Stone,[1] Benjamin Whitefield,[2,3] Mehran Kianinia[2,3] and Carlo Bradac[1, *]

[1] *Trent University, Department of Physics & Astronomy, 1600 West Bank Drive, Peterborough, ON, K9L 0G2, Canada.*
[2] *School of Mathematical and Physical Sciences, University of Technology Sydney, Ultimo, NSW 2007, Australia*
[3] *ARC Centre of Excellence for Transformative Meta-Optical Systems, University of Technology Sydney, Ultimo, NSW 2007, Australia*

* corresponding author: Carlo Bradac, carlobradac@trentu.ca

**Abstract**
Optically detected magnetic resonance (ODMR) has become a well-established and powerful technique for measuring the spin state of solid-state quantum emitters, at room temperature. Relying on spin-dependent recombination processes involving the emitters' ground, excited and metastable states, ODMR is enabling spin-based quantum sensing of nanoscale electric and magnetic fields, temperature, strain and pressure, as well as imaging of individual electron and nuclear spins. Central to many of these sensing applications is the ability to reliably analyze ODMR data, as the resonance frequencies in these spectra map directly onto target physical quantities acting on the spin sensor. However, this can be onerous, as relatively long integration times—from milliseconds up to tens of seconds—are often needed to reach a signal-to-noise level suitable to determine said resonances using traditional fitting methods. Here, we present an algorithm based on data clustering that overcome this limitation and allows determining the resonance frequencies of ODMR spectra with better accuracy (∼1.3 × factor), higher resolution (∼4.7 × factor) and/or overall fewer data points (∼5 × factor) than standard approaches based on statistical inference. The proposed clustering algorithm (CA) is thus a powerful tool for many ODMR-based quantum sensing applications, especially when dealing with noisy and scarce data sets.

## 1. Introduction

Spin defects in solid-state host materials have become prime candidate hardware systems for advanced quantum technologies.[1–4] Prominent representatives include atom-like quantum emitters in diamond,[5,6] silicon carbide,[7] hexagonal boron nitride (hBN),[8,9] zinc oxide,[10] transition metal dichalcogenides,[11] as well as rare-earth ions in solids,[12,13] and optically-active donors in silicon.[14,15] These atom-like emitters are attractive for quantum applications as they often possess spin states that can be readily manipulated and read out, whilst also displaying long coherence time, room-temperature operation, the ability to create entangled states, and spin-dependent optical transitions that allow for spin-photon interfacing and long-distance transmission of quantum information.[16,17] Additionally, the solid-state host materials are technology-ready. Potential devices can be realized leveraging well-established nanofabrication techniques from the semiconductor industry and hold prospects for seamless integration with on-chip electronic, magnetic and photonic nanostructures.[18–20] Relevant applications span over a wide range of fields including quantum communication[21–24] and computation,[25–28] quantum simulation,[29,30] and quantum metrology and sensing.[4,31,32]

In this work, we focus primarily on quantum sensing—specifically on the ability to map external stimuli onto frequency shifts in the optically detected magnetic resonance (ODMR) spectra of solid-state spin defects. Realized with quantum emitters in diamond,[33] silicon carbide (SiC),[34] hexagonal boron nitride (hBN)[32] and most recently gallium nitride (GaN),[35] ODMR has become a well-established and powerful technique with applications that include measuring electric and magnetic fields, temperature, strain and pressure, as well as imaging electron and nuclear spins[36–38]—all with remarkably high sensitivity and nanoscale spatial resolution. Notably, in the practical measurement of any of these physical quantities, resolution and accuracy depend on the ability to identify the resonance frequencies in the ODMR spectrum. Therefore, long integration times are generally desirable in ODMR measurements as they yield better photoluminescence (PL) contrast and signal-to-noise ratio. This, however, inevitably leads to lengthy acquisitions, as obtaining a single ODMR spectrum can require scanning the microwave (MW) signal over a relevant range of frequencies (~hundreds of MHz) while recording the PL contrast with dwell times of the order of $\sim 10^{-3}$–$10^2$ s, at every single one value of MW frequency. Here, we demonstrate a method based on data clustering that can extract the resonance frequencies from ODMR spectra with better accuracy (~1.3 × factor) and higher resolution (~4.7 × factor) than traditional methods based on statistical inference. Equivalently, this means that for specified levels of accuracy and resolution, our method can determine the resonance frequencies of ODMR data with a ~5 × factor fewer data points than traditional fitting methods, effectively speeding up and making more efficient both the ODMR acquisition and interpretation processes.

## 2. Materials and Methods
### 2.1 Spin-based sensing: theoretical framework
Optically detected magnetic resonance (ODMR) occurs when a quantum emitter displays photoluminescence (PL) that depends on the electron spin state of the system (figure 1a). That is, under optical excitation, the system emits an average number of photons determined by the specific spin-state-dependent transitions—radiative or non-radiative—the emitter undergoes. The mechanism relies on the system displaying paramagnetic behavior (total spin angular momentum $S \geq 1$) in either the ground and excited states or the metastable states.[5,39,35] Figure 1b shows two model level diagrams for atom-like systems displaying ODMR. The diagram to the left is representative of systems such as the nitrogen-vacancy (NV) center in diamond and the boron-vacancy ($V_B^-$) center in hexagonal boron-nitride, which can be taken as archetypes—analogous considerations can be made for other spin defects in solid-state hosts. For instance, the NV center has both a triplet ($S = 1$) ground and excited state, and singlet ($S = 0$) metastable state(s). The transitions between the ground and excited triplets are mostly spin conserving ($\Delta m_s = 0$) and are associated with the absorption and emission of photons. Conversely, the intersystem crossing (ISC) transitions from the excited sublevels to the singlet metatable state(s) are non-radiative and spin selective—specifically, the shelving rate from the $m_s = 0$ excited state sublevel is lower than those from the $m_s = \pm 1$ ones. From the singlet metastable state(s) the emitter decays preferentially towards the ground state $m_s = 0$ sublevel. These spin-selective processes result in both a non-Boltzmann steady-state alignment of the electron spin into the $m_s = 0$ sublevel through optical pumping and, since ISC transitions are non-radiative, to a higher detected PL intensity when the $m_s = 0$ sublevel is preferentially populated over the $m_s = \pm 1$ ones. The relative population of the $m_s = 0$ and $m_s = \pm 1$ states can be controlled through application of a MW field resonant with the corresponding transition between sublevels, which results in the characteristic PL contrast signature peaks observed in ODMR spectra.[40] Determining the exact position of these frequency resonances is at the core of many ODMR-based quantum sensing applications as their relative position is denotative of the surrounding environment and presence of external stimuli such as electric and magnetic fields, strain, etc. Taking again the NV center as an archetype system, the Hamiltonian of its ground state triplet can be written as:[41]

$$H_{gs} = \left(hD_{gs} + d_{gs}^{\parallel}\Pi_z\right)\left[S_z^2 - \frac{1}{3}S(S+1)\right] + \mu_B g_e \boldsymbol{S} \cdot \boldsymbol{B} - d_{gs}^{\perp}\left[\Pi_x\left(S_x S_y + S_y S_x\right) + \Pi_y\left(S_x^2 - S_y^2\right)\right] \quad (1)$$

Where $d_{gs}^{\parallel}$ and $d_{gs}^{\perp}$ are the axial and non-axial components of the ground triplet state permanent electric dipole moment $d_{gs}$, respectively, $h$ is the Planck constant, $\mu_B$ is the Bohr magneton, $g_e$ is the electron g-factor, $\boldsymbol{B}$ is the applied magnetic field, and $\boldsymbol{S}$ is the electron spin operator. Note that in equation (1) the coordinate system is defined such that the z-axis is parallel to the N-V axis of the center. Equivalent, specific expressions can be written for the ground, excited and/or metastable states of any other solid-state spin defect.

Relevant for quantum sensing, is the fact that in the Hamiltonian of equation (1), the effect of local strain, $\boldsymbol{\sigma}$, external electric fields, $\boldsymbol{E}$, (lattice strain can be treated as a local static electric field such that $\boldsymbol{\Pi} = \boldsymbol{E} + \boldsymbol{\sigma}$ in (1)[42]) magnetic fields, $\boldsymbol{B}$, as well as temperature, $T$, can be mapped directly as changes in the relative energies between sublevels $m_s = 0 \leftrightarrow \pm 1$, which correspond to the resonant frequencies $f_1$ and $f_2$ measured in ODMR experiments. This is evident in figure 1c, which shows the ODMR resonant frequencies of a diamond NV center shifting under the effect of a changing magnetic field.

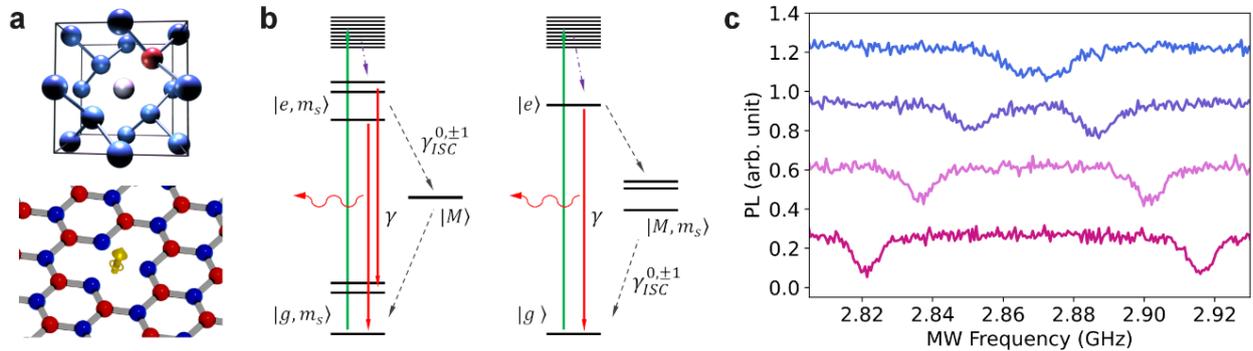

**Figure 1.** Principles of ODMR-based sensing. **a)** Crystalline structure of two common spin-defects in solid-state hosts: the nitrogen-vacancy (NV) center in diamond (top) and of the boron-vacancy ($V_B^-$) center in hexagonal boron-nitride (bottom). **b)** Archetypal level diagrams of spin-defects displaying $S \geq 1$ ground/excited states (left) and $S \geq 1$ metastable state(s) (right). The non-radiative intersystem crossing rates, $\gamma_{ISC}^0$ and $\gamma_{ISC}^{\pm 1}$, are spin-dependent and lead to the characteristic resonance frequencies observed in the ODMR spectrum when specific sublevels are populated (see main text). **c)** Room temperature ODMR spectra of a single diamond NV center under the effect of a magnetic field of increasing magnitude. Application of a magnetic field, $\boldsymbol{B}$, causes a shift (Zeeman effect) in the relative energies of the triplet (ground) spin sublevels of the center (equation 1). The relative energies are detected as optical contrast in the photoluminescence (PL) signal occurring when the driving microwave field is resonant with the corresponding transitions between spin sublevels.

In practical measurements, it is important to be able to isolate the different effects of any one physical quantity, as $E, B, \sigma, T,$ etc. all contribute to the energy of the system as per equation (1). Also relevant for practical sensing applications—and of primary interest in this work—is the achievable resolution of the measurement. The resolution of any spin-based sensor is ultimately limited by the quantum noise associated with spin-projection, but it additionally depends, when the detection is optical, on photon shot noise, quantum efficiency of the emitter, as well as photon extraction/collection efficiency of the measurement system. Depending on the specificity of the measurement, phenomena affecting linewidth and spin coherence time can also play a role in limiting the resolution of the quantum sensor.[33] In light of these considerations, our focus in this work is to demonstrate a strategy to measure the resonance frequencies in ODMR spectra accurately and efficiently—especially when the data sets are noisy and/or scarce.

*2.2 Data Analysis*

The algorithm we developed to analyze ODMR spectra is based on data clustering (§2.3). To test the performance of the algorithm we use both experimental and synthetic/emulated ODMR data of boron-vacancy ($V_B^-$) centers in hexagonal boron nitride (hBN), which are spin-defects with $S \geq 1$ ground/excited states (figure 1b, left).

Experimental data was collected from ensembles of $V_B^-$ centers in hexagonal boron-nitride. A flake of hexagonal boron nitride (hBN) was first exfoliated onto a $SiO_2$ substrate and irradiated with nitrogen ions at a fluence of $10^{14}$ ions/cm$^2$ to create ensembles of negatively charged boron vacancy ($V_B^-$) centers. A microwave (MW) antenna was patterned using UV photolithography, initially sputtering 5 nm of chromium as a sticking layer followed by 500 nm of gold to form the antenna itself. The nitrogen irradiated hBN flake was align-transferred on top of the MW antenna using a polyvinyl alcohol (PVA) coated polydimethylsiloxane (PDMS) stamp.

Optical excitation of the $V_B^-$ ensembles was performed using a 532-nm laser on a lab-built confocal microscope.[43] An air objective (numerical aperture, $NA = 0.9$, and magnification 100 ×) was used for excitation and back collection. A 532-nm dichroic mirror, as well as a 568-nm long pass filter were employed to separate the laser excitation from the $V_B^-$ emission (~800 nm). The emission was then coupled into an optical fiber and detected by a single-photon avalanche photodetector (APD).

Optically detected magnetic resonance (ODMR) was performed on the $V_B^-$ ensembles via constant-wave (CW) excitation while sweeping a MW signal from 3 to 4 GHz. Each frequency step involved activating the MW signal for 1 ms followed by a 1 ms period with no MW signal, serving as a reference measurement. The emitted photons were then collected by the APD and the reference signal was used to calculate the ODMR contrast. An external magnetic field was also applied from below the sample to frequency-shift the $m_s = \pm 1$ spin resonances. Four different magnetic field strengths were applied to the sample and five excitation locations were measured for each magnetic field strength. For each measurement, the frequency sweeps are saved individually and are integrated to obtain total acquisition times. All measurements were carried out using 5 mW of laser power (measured at the back aperture of the objective).

Additionally, synthetic or emulated data sets were generated by simulating ODMR experiments in which the values for the resonance frequencies, their full width at half maximum (FWHM) and photoluminescence contrast are assumed to follow Lorentzian distributions and are known. We remark that the choice of the Lorentzian distribution for the simulated ODMR peaks is not a strict requirement; in fact, one of the advantages of the proposed algorithm is that it is agnostic to the functional shape and would perform equally well with data following, e.g., Gaussian or Voigt distributions (which might occur in samples where emission shows inhomogeneous broadening). The motivation for choosing the Lorentzian function is twofold. Firstly, our experimental ODMR spectra for $V_B^-$ centers in hBN follow this distribution. Secondly, this choice allows us to carry out a direct and quantitative comparison between the performance of our algorithm and that of standard fitting methods. Note that in generating the synthetic data we keep the constraints at a minimum: the resonance frequencies are set to occur anywhere within the 3–4 GHz range and their FWHM and PL contrast are independent and are let to vary freely and randomly in the range 30–135 MHz and 0.03–0.1, respectively. In real quantum systems some of these parameters have stronger constraints dictated by the physics of the system itself (e.g. there is an inherent symmetry for the transitions $m_s = 0 \leftrightarrow +1$ and $m_s = 0 \leftrightarrow -1$). However, in our simulated data we deliberately relax these constraints.

This is done in favor of generality, but also to account for measurement artifacts that we observe in the experimental ODMR data that are due to our MW amplifier displaying a non-uniform, frequency-dependent response (see discussion in §2.3 and §3). By simulating the ODMR experiments we can set every relevant parameter with arbitrary precision. Additionally, since the synthetic data sets are known with absolute certainty, we can directly quantify the accuracy and resolution of any data analysis technique and specifically benchmark our algorithm against traditional methods based on statistical inference, e.g. Levenberg–Marquardt (L-M) or least-squares fit.

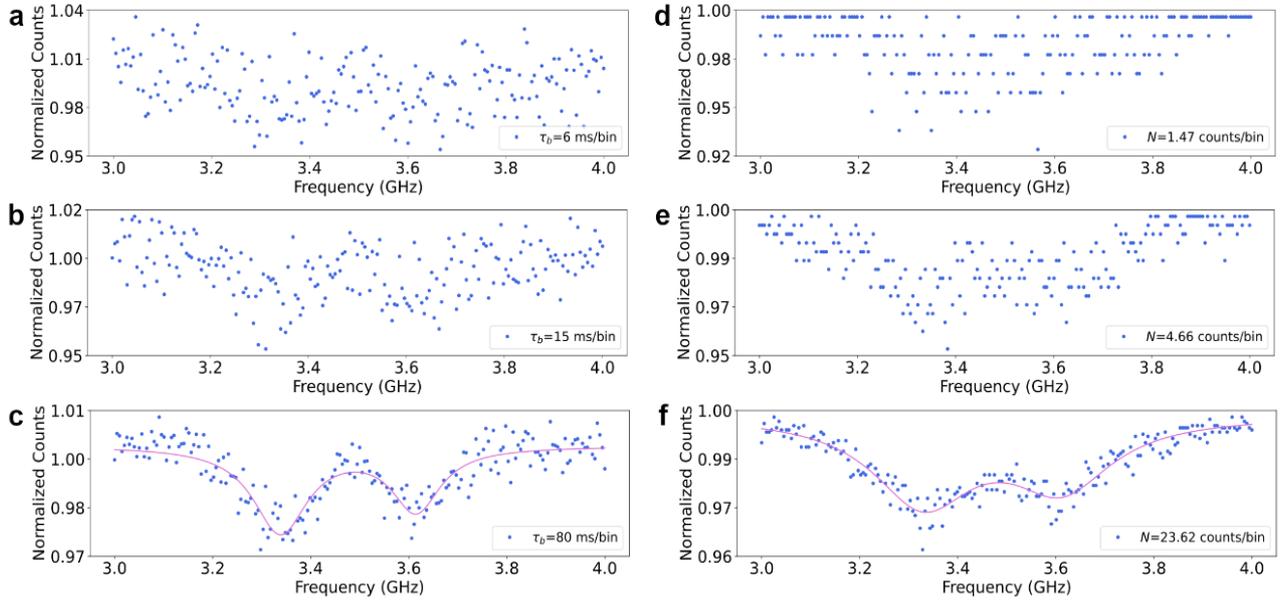

**Figure 2.** Experimental and synthetic ODMR spectra. **a–c)** Experimental ODMR data obtained from ensembles of $V_B^-$ centers in hBN. The spectra were acquired integrating the signal for times, $\tau_b$, equal to $6\ \mathrm{ms/bin}$ (a), $15\ \mathrm{ms/bin}$ (b) and $80\ \mathrm{ms/bin}$ (c); these correspond to total integration times of $1.2\ \mathrm{s}$, $3.0\ \mathrm{s}$ and $16\ \mathrm{s}$, respectively, as the microwave signal is swept over the range $3–4\ \mathrm{GHz}$ in $200$ steps at $5\ \mathrm{MHz}$ increments. **d–f)** Synthetic ODMR data. The simulated spectra have an average number of counts per bin, $N$, equal to $1.47\ \mathrm{counts/bin}$ (d), $4.66\ \mathrm{counts/bin}$ (e) and $23.62\ \mathrm{counts/bin}$ (f). The continuous lines in (c) and (f) are Lorentzian fits to the data.

For our analysis, we use up to $10^5$ simulated data sets per value of $N$ (defined below) generated on a supercomputer; we also test the performance of the algorithm on real data sets. Figures 2a–f show a few examples of both real and simulated ODMR spectra with different values of (photo)counts. Real ODMR data sets are labeled using the integration time per bin, $\tau_b$, employed in the measurement, while synthetic data sets are labeled based on the average number of counts per bin, $N$. These two quantities can be related directly to one another as the number of counts per bin in real data correlates directly to the integration time through the detection efficiency of the setup. In our case, $1\ \mathrm{ms/bin}$ integration time of experimental data corresponds to $\sim 10^7\ \mathrm{counts/bin}$. The absolute average number of counts per bin is not important; what matters is the relative number of tallied counts for the frequency bins that are 'off-resonance' vs. those that are 'on-resonance,' as the latter determine the ODMR resonances we are interested in.

Whilst for the spectra in figure 2c and 2f finding the resonant frequencies is straightforward using traditional fitting methods, these methods can be quite inaccurate for scarce data sets such as those of figure 2a, b, d and e, where they are often unable to converge. Yet, being able to extract the correct value of the resonance frequencies in these latter cases is desirable as it reduces the overall measurement time and allows dealing with low contrasts and noisy data sets. These aspects warrant two relevant observations. Firstly, regarding the measurement time, it is worth emphasizing that traditional ODMR spectra are often acquired by sweeping the MW field in a given range at fixed incremental steps while photocounts are measured during a specified dwell time, at each one of these steps. For instance, in our experiments we scan the MW field over the range 3–4 GHz in 200 steps at 5 MHz increments. As a result, even relatively short dwell times (tens of ms) at each value of frequency, can lead to total integration times that are lengthy (~seconds to tens of seconds). Secondly, determining the resonance peaks can be onerous when low contrast in the ODMR acquisition (determined by the photo-physics of the emitter via the rates γ$_{ISC}$) is combined with low signal to noise ratio.[35,44] This matters as uncertainty in determining the resonance frequencies directly translates in lower resolutions of spin-based sensing techniques (cf. §3).

*2.3 Data clustering algorithm*

The algorithm we developed to analyze ODMR spectra is a custom-written algorithm that uses K-means clustering[45,46] and the open-source machine learning library scikit-learn[47] as foundation. The algorithm involves two main stages of clustering and includes a set of conditions on the clusters to handle various cases. Briefly, the data is first clustered 'vertically,' as the 1D array for the number of PL counts is divided in clusters that are referred to as *rows*. The number of these vertical clusters ($k_v$) is determined by a custom, quantitative version of the *elbow method*, where the so-called inertia (i.e., the sum of squared distances of samples to their closest cluster center, weighted by the sample weights)[47] is plotted and evaluated against the number of clusters. The optimal number of clusters is determined at the 'elbow point,' i.e., where the rate of *explained variance* levels off, which in our data sets consistently occurs for $k_v = 4$ rows. After this first 'vertical' clustering, the data undergoes a second 'horizontal' clustering step. Specifically, data points from the lowest of the $k_v$ clusters are clustered into $k_h = 2$ 'horizontal' groups of 1D microwave frequency data, which we refer to as *columns*. The centroid of these two horizontal clusters or columns are taken as the values of the resonance frequencies in the ODMR spectrum. Figure 3a illustrates, visually, the vertical and horizontal clustering steps applied to an ODMR spectrum where the resonant frequency peaks are obvious and easily discernible.

Our algorithm also includes a few features to directly tackle an issue that affects some of the experimental ODMR spectra, shown in figure 3b. In some of the spectra, there is a significant difference (up to 65%) between the PL contrast values of the two resonance peaks. This is due to the microwave amplifier not having a uniform power output across the spanned frequency range, 3–4 GHz—the power output is effectively lower at higher frequencies. If the contrast values of the PL peaks in a spectrum are very different, the first step of vertical clustering might miss the cluster for the peak with the lower contrast (figure 3b). The algorithm accounts for this by performing crosschecks between the horizontal *columns* of each of the vertical *rows*. It specifically compares the distance between the column's centroids, widths, and the average distance between the maximum and minimum points in different rows as well as in groups of rows (see Supplementary Information for details). These crosschecks also allow the algorithm to handle cases where the two $m_s = \pm 1$ sublevels are degenerate in energy, resulting in the two resonant frequencies overlapping in the ODMR spectra (figure 3c).

**3. Results and discussion**

To characterize the performance of our algorithm, we determine its accuracy and precision, and benchmark them against those of traditional fitting methods based on statistical inference. Accuracy is measured as the difference between the values of the resonance frequencies returned by the algorithm and their 'true' values; precision is measured as the standard deviation of these returned values, determined from all the data sets we analyzed ($10^5$ spectra for each value of $N$ in figures 3d, e). Note that precision translates directly into resolution for any spin sensing technique that maps external physical quantities ($E$, $B$, $\sigma$, $T$, etc.) onto the relative positions of the ODMR resonance frequencies, as it dictates the minimum resolvable difference one can measure.[48] In our characterizations, the 'true' values of the resonance frequencies are either their values set in the simulated ODMR spectra (in which case they are exact and known with absolute certainty) or the values determined from ODMR experiments carried out with long integration time, $\tau_b = 10 \text{ s/bin}$, and high signal to noise ratio, $\sim 500$.

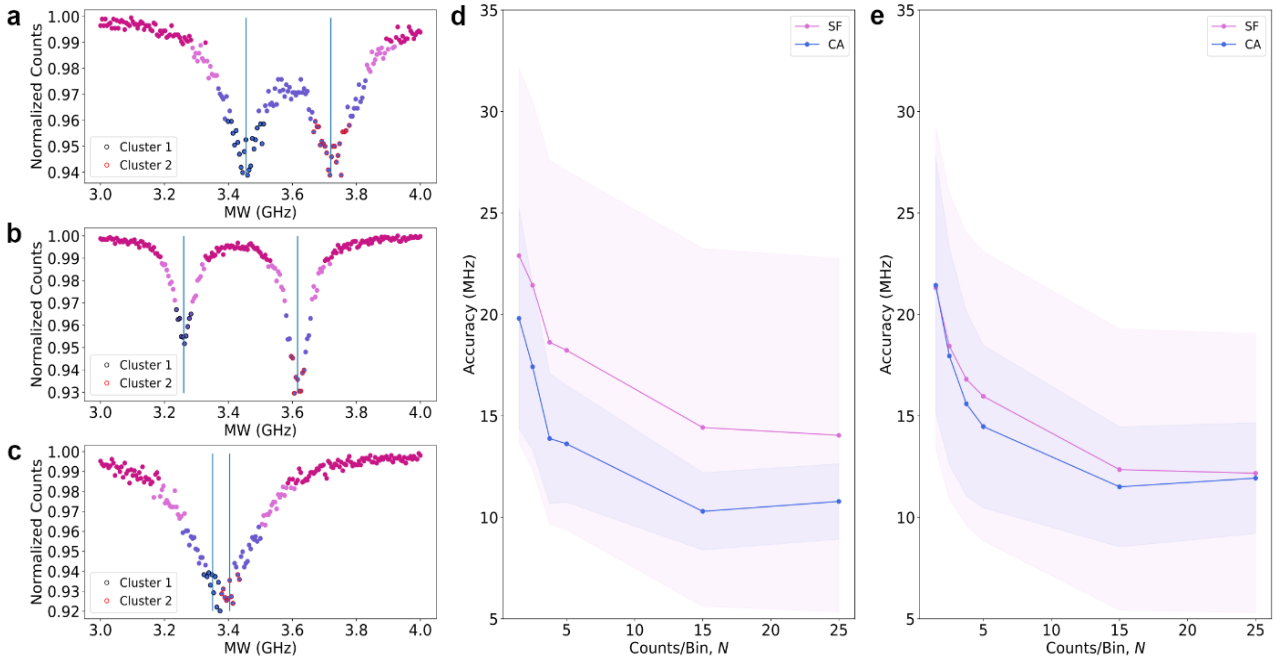

**Figure 3.** Clustering algorithm: operation and performance. **a–c)** Graphic illustration of the working principle of the clustering algorithm (CA) which assigns data to vertical (*rows*) and horizontal (*columns*) clusters. The process outputs two final *clusters* (labeled 'Cluster 1' and 'Cluster 2') that correspond to the resonance frequencies of the ODMR spectra. Three representative cases are displayed, showing ODMR spectra where the resonant peaks have similar (a) and different (b) PL contrast values, and where the peaks are overlapping (c). **d, e)** Accuracy of the clustering algorithm (CA, dark blue) and standard fitting (SF, light pink) methods plotted vs. the average number of counts/bin, $N$, used in the simulated data. Lower values indicate better accuracy (i.e., smaller distance from the 'true' value). The precision of each method is indicated by the corresponding amplitude of the shaded area; for representation purposes, the amplitudes of the shading have been divided by a factor five for all sets in both (d) and (e). Smaller amplitudes indicate higher precision (i.e., smaller error). The graphs in (d) and (e) differ as they are obtained assuming the largest difference in PL contrast between the two ODMR peaks to be as high as $15\%$ or $65\%$, respectively.

The results of our analysis are shown in figures 3d and 3e. In both graphs, the accuracy for the values of the resonance frequencies is plotted against the average number of counts/bin, $N$. The graphs also display the precision, which is given by the amplitude of the shading centered around each point (for visualization purposes the amplitude of the shading has been divided by a factor five both in figures 3d and 3e). In the plots, lower values of the points and smaller amplitudes for the shaded areas are desirable as they correspond to better accuracy (smaller distance from the true values) and better precision/resolution (smaller standard deviation or error), respectively. The dark blue sets show the results for our clustering algorithm (CA), while the light pink sets show those for standard fitting (SF) methods. Note that the plots in figures 3d and 3e are homologous: they only differ from one another based on the simulated ODMR data we used. The values for figure 3d were determined assuming that the PL contrast values in the resonance frequencies of the simulated ODMR spectra are the same (difference $\leq 15\%$). Conversely, the values for figure 3e were determined allowing a difference in PL contrast between peaks $\leq 65\%$ (which is the largest difference we measured in our experimental ODMR data, due to the non-uniform power output of our MW amplifier). Regardless of the sets we consider, the clustering algorithm performs better than traditional fitting methods and displays overall better accuracy and precision: the data points for the CA are lower on the graph than those for SF and they have smaller shading/uncertainty. For instance, looking at the plots in figure 3d, the clustering algorithm achieves similar, or better, values of accuracy than standard fitting methods with much fewer (factor $\geq 5 \times$) average counts/bin, $N$, and whilst also maintaining an overall higher resolution (factor $\geq 3 \times$).

These results call for some observations. In general, the direct comparison between methods in figures 3d and 3e highlight that the proposed clustering algorithm (CA) can indeed be a powerful tool for speeding up the ODMR acquisition and data analysis processes, with improvements in efficiency by a factor $\geq 5 \times$. We underscore that when the average number of counts/bin and the signal-to-noise ratio for the ODMR data increase (at longer simulation/integration times), the CA and SF methods converge in performance and reach the same levels of accuracy and precision. However, the CA algorithm clearly outperforms the SF method for data sets that are noisy and/or scarce (at short simulation/integration times)—which is attractive, as ODMR experimental measurements can be relatively lengthy (see final observation below).

Additionally, note that the CA is an *unsupervised* algorithm, meaning that it does not require training with priorly-acquired and known sets of data. This is desirable as training of the algorithm in *supervised* methods requires time and computing resources, as well as access to training and test data sets that are usually large and lengthy to acquire.

The CA algorithm is model-agnostic: that is, no assumptions are made about the functional form of the ODMR resonance peaks, e.g., Lorentzian, Gaussian, Voigt, etc.—which is instead required in traditional fitting approaches.

Finally, by design, the CA always returns values for the resonance frequencies, even for exceptionally scarce data sets. For demonstration purposes, we tested the CA over $10^5$ sets of emulated ODMR data with an average number of counts per bin as low as $N = 0.05$. As expected, the values for accuracy and resolution are undesirably large—$89\,\text{MHz}$ and $106\,\text{MHz}$, respectively. Yet this information is still valuable, as it can be used, for instance, to implement dynamic-type ODMR measurements where instead of scanning the MW signal 'indiscriminately' and uniformly across a preset range of frequencies ($3–4\,\text{GHz}$ in our experiment), the MW is swept over smaller and targeted frequency windows, informed by the values returned by the CA. This is not practically feasible using standard fitting methods that, for comparison, using $N = 0.05$ fail to converge at all in over $60\%$ of the tested ODMR data sets.

## 4. Conclusions

To conclude, we have presented an algorithm to efficiently analyze optically detected magnetic resonance (ODMR) data from quantum emitters. The algorithm is a custom variant based on data clustering and inspired by the K-means method. Through the employment of both synthetic and experimental data, we demonstrate that the algorithms can extract the resonance frequencies of ODMR spectra with better accuracy ($\sim 1.3 \times$ factor), higher resolution ($\sim 4.7 \times$ factor) and/or overall fewer data points ($\sim 5 \times$ factor) than traditional approaches based on statistical inference, such as Levenberg-Marquardt and least-squares fitting methods. Our clustering algorithm (CA) is therefore a powerful tool for improving the accuracy, speed and efficiency of any quantum sensing application based on detecting the resonance frequencies of spin defects from ODMR measurements, especially those involving noisy and scarce data sets.

**5. Data Availability**

The Supplementary Information file contains additional details about the working principles of the clustering algorithm. The synthetic ODMR data and code of the CA used in this study are available in a public repository.
https://github.com/DylanStone2000/Fast-characterization-of-optically-detected-magnetic-resonance-spectra-via-data-clustering

**5. Contributions and Acknowledgements**

Experimental ODMR data was collected by B.W. and M.K. Theoretical modeling, simulation and data analysis was performed by D.G.S. and C.B. All authors contributed to writing the manuscript. The authors declare that they have no conflicts of interest or competing interests related to the research presented in this paper.

The Natural Sciences and Engineering Research Council of Canada (via GPIN-2021-03059 and DGECR-2021-00234) and the Canada Foundation for Innovation (via John R. Evans Leaders Fund #41173) are gratefully acknowledged. The authors also acknowledge the Australian Research Council (CE200100010, FT220100053, and FT200100073) and the Office of Naval Research Global (N62909-22-1-2028) for the financial support.

# Supplementary Information:
## Fast characterization of optically detected magnetic resonance spectra via data clustering

Dylan G. Stone,[1] Benjamin Whitefield,[2,3] Mehran Kianinia,[2,3] and Carlo Bradac[1, *]

[1] Trent University, Department of Physics & Astronomy, 1600 West Bank Drive, Peterborough, ON, K9L 0G2, Canada.
[2] School of Mathematical and Physical Sciences, University of Technology Sydney, Ultimo, NSW 2007, Australia
[3] ARC Centre of Excellence for Transformative Meta-Optical Systems, University of Technology Sydney, Ultimo, NSW 2007, Australia
*Corresponding author, e-mail: carlobradac@trentu.ca

**Data clustering algorithm**

As discussed in the manuscript, our clustering algorithm (CA) involves two main clustering *steps*:

1) 'Verticallly' clustering the 1D array containing photoluminescence (PL) counts into *rows*.
2) 'Horizontally' clustering the 1D array containing the microwave (MW) frequency data into two *columns*.

The algorithm contains a set of *conditions* that help the second step correctly identify the data that should be included in the analysis, depending on the characteristics of the spectrum. Here we explain in detail the working principles of the CA and these additional conditions. We discuss three sets of ODMR data that are taken as representative 'archetypes' covering most possible scenarios.

The three types of data explored in this study were: **a)** clearly separated peaks with a difference in PL contrast of up to 65% ('asymmetric' data), **b)** clearly separated peaks with a difference in PL contrast of up to 15% ('symmetric' data), and **c)** overlapping peaks. These three cases are shown in figure S1.

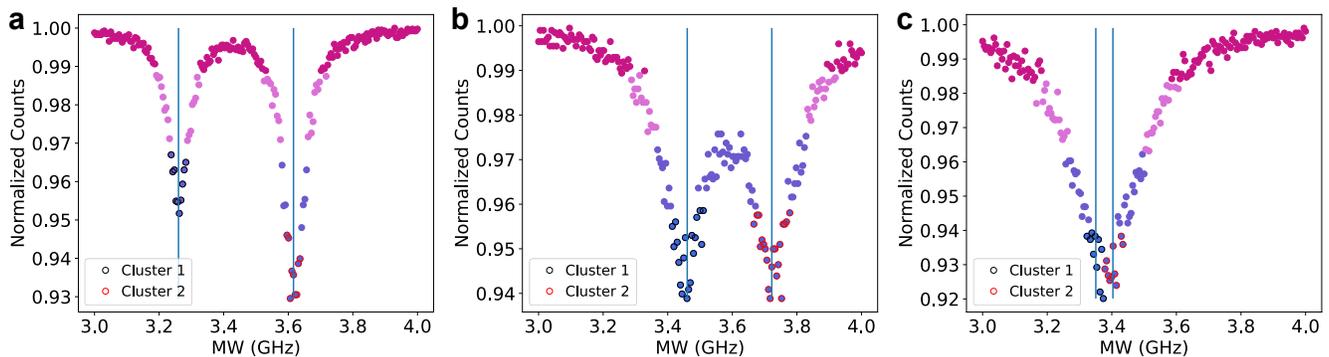

**Figure S1**. Examples of the three possible ODMR data archetypes considered in this analysis. **a)** PL contrast of up to 65% (asymmetric). **b)** PL contrast of up to 15% (symmetric). **c)** partial-to-fully overlapping peaks.

To differentiate between these three instances, two *conditions* were imposed to check the structure of the data. *Condition 1* looks at the changes in the column width (in this case the average of both column widths) compared to the changes in the column distances (the distance between cluster centroids) when in the second 'horizontal' clustering step we include *i)* only the lowest row, or *ii)* the lowest two rows. This condition was designed specifically to identify the cases like that of figure S1a, where just using the lowest row of data in our second step would completely miss the left peak. The idea is that if both peaks were present in the lowest row, we expect that including the second lowest row should increase the width of the columns as well as the distance between them, since the peaks widen asymmetrically. Conversely, if one of the peaks was not present, we would expect a larger change in the column distance compared to the column width by including the second lowest row. This is shown in figure S2.

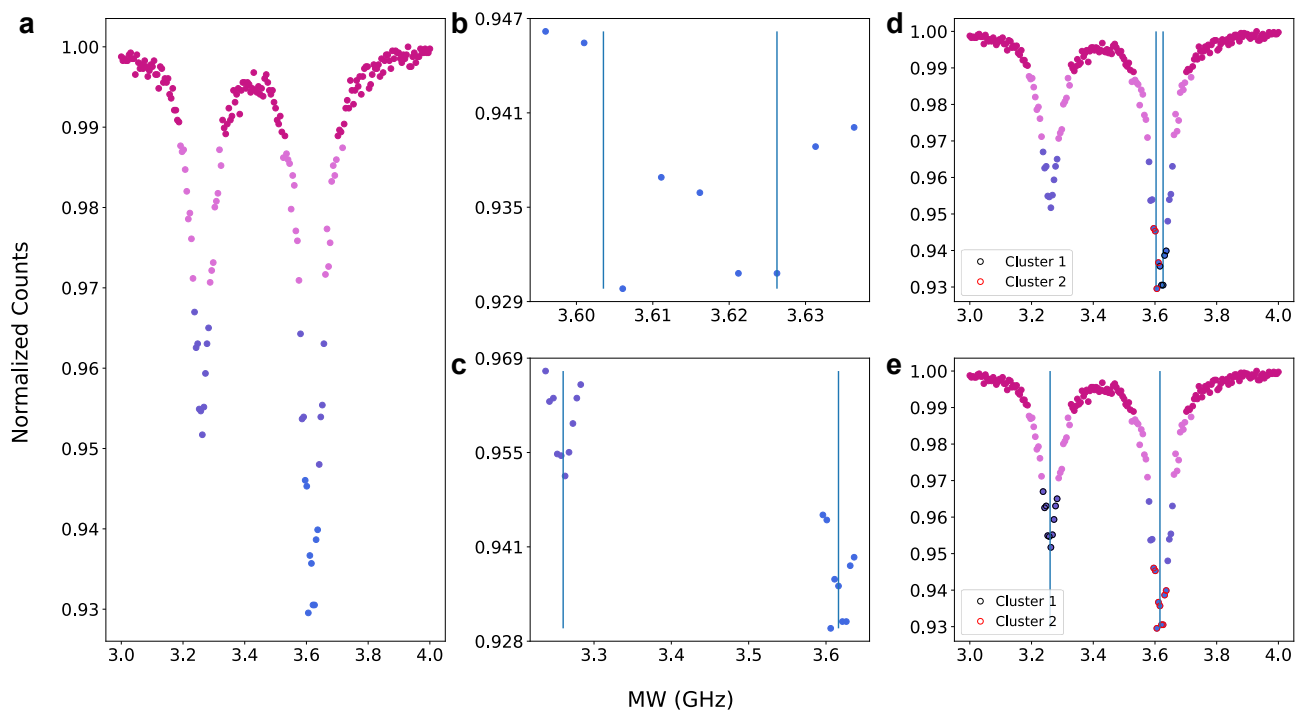

**Figure S2**. Visualization of the process to select the correct peaks. **a)** The raw data is color coded; each color indicates a different row as identified by the first clustering step. **b-c)** Subset of data that is passed to step two not using (b) and using (c) the *conditions*, respectively. **d-e)** Resulting peak prediction based on not using (d) and using (e) the *conditions*, respectively.

For example, in figure S2, since only one peak located at ~3.62 GHz is present in the lowest row, the two columns identified by the algorithm have centroids close to 3.62 GHz (~3.60 and ~3.63 GHz) as shown in figure S2b. However, also including the second row reveals a second peak at 3.26 GHz, as shown in figure S2c. This results in the distance between the columns dramatically increasing compared to the column widths, as the centroids are at ~3.26 and ~3.62 GHz, with comparatively slight increases in the column widths. With this in mind, we can set *Condition 1*:

$$\frac{next\ distance/current\ distance}{next\ width/current\ width} \leq 1$$

to determine if we should include the second lowest row. If this ratio is less than or equal to one, then including the next row does not reveal a previously missed peak (Note: the model currently identifies missed peaks one row above, the code can be easily modified to check any number of rows above for missing peaks if the data contains extremely large PL contrasts, at the cost of run time). The specific choice of 1 for the threshold of the ratio in *condition 1* was found to work well with the artifact in our experimental data. The value was chosen by comparing the ratio of distance-to-width change for datasets that were known to miss peaks to those that always identified both.

However, including unnecessary rows can have a negative impact on the accuracy of the peak predictions, due to the asymmetric broadening of the peaks as more rows are included, which can shift the centroid away from the 'true' peak position. To avoid using unnecessary points in the second step of clustering, we included a *Condition 2*, which checks if either of the previous columns are contained within either of the new columns. In practice, it is sufficient to only check whether one of the previous columns is contained within either of the new columns. For leniency, *Condition 2* checks if at least 80% of the data points are present:

$$\{Previous\ columns\} \subset^{0.8} \{Current\ columns\}$$

Here we use $\subset^{0.8}$ to denote a fractional subset. If this condition is met, the column that contains the previous columns is instead replaced with them, such that it only contains data from the lowest row. In figure S2c) the blue cluster on the right would have also contained the purple points from the above row, but since the blue points in the lowest row (the previous two columns as seen in figure S2b) are within the new column containing both rows, the upper row is excluded. We are now only including the second lowest row for the missed peak in the second step of clustering.

In the rare instance that the second row is used, but the second condition is not met, we are dealing with our third situation (figure S1c) where the peaks are nearly or fully overlapping. Due to the asymmetric broadening of the peak(s), as we move up the rows the location in which the clump of points is divided can shift, causing neither of the previous columns to be fully contained within either of the new columns which contain the lowest two rows. In this case, the algorithm automatically reverts back to using only the lowest row. However, as mentioned this is an uncommon occurrence: typically, overlapping peaks will satisfy the *condition 1*, skipping over the *condition 2* entirely.

It is important to note that in cases like that of figure S1c, due to the way the second set of clustering occurs, the algorithm always returns two values for the predicted peaks even if they were completely overlapping. In our CA, this becomes a source of uncertainty (loss of accuracy and precision) in determining the resonance frequencies. Despite this, the model still outperforms the standard fit in our testing.